\documentclass[lettersize,journal]{IEEEtran}
\usepackage{amsmath,amsfonts}
\usepackage{algorithmic}
\usepackage{algorithm}
\usepackage{array}
\usepackage{textcomp}
\usepackage{stfloats}
\usepackage{url}
\usepackage{verbatim}
\usepackage{graphicx}
\usepackage{cite}
\usepackage{color,soul}
\usepackage{subcaption}
\begin{document}

\title{Data-driven surrogate model for etch rate profiles using sensor data from a reactive ion etcher}

\author{Abhijit Pranav Pamarty, Robert Neuweiler, Le Quyen Do, Keaton Johnson, James J. Sanchez, Dinesh Koli}


\maketitle

\begin{abstract}
Reliable predictions of the etch rate profile are desirable in semiconductor manufacturing to prevent etch rate target misses and yield rate excursions. Conventional methods for analyzing etch rate require extensive metrology, which adds considerable costs to manufacturing. We demonstrate a data driven method to predict the etch rate profiles of a capacitively-coupled plasma RIE etcher from the tool’s sensor data. The model employs a hybrid autoencoder-multiquadric interpolation-based approach, with the autoencoder being used to encode the features of the wafers’ etch rate profiles into a latent space representation. The tool’s sensor data is then used to construct interpolation maps for the latent space variables using multiquadric radial basis functions, which are then used to generate synthetic wafer etch rate profiles using the decoder.  The accuracy of the model is determined using experimental data, and the errors are analyzed in interpolation and extrapolation.
\end{abstract}

\begin{IEEEkeywords}
Dry etch, autoencoders, etch rate modeling, process optimization, convolutional neural networks, surrogate modeling 
\end{IEEEkeywords}

\section{\label{sec:level1} Introduction}

Reactive ion etching is a microfabrication technique in which a plasma is used to remove material through the combination of physical bombardment and chemical etching\cite{jansen1996survey, oehrlein1986reactive}. A strong alternating RF electromagnetic frequency is used to strip away the electrons from a gas, which then oscillate strongly with the applied electric field and are absorbed into the walls of the chamber. This then leaves a positively-charged ionic plasma that etches the surface of the wafers \cite{hinson1984low, oehrlein1989competitive}.  

Understanding of the process's etch rate allows for more controllable process conditions, which may better match tool performance to the requirements of a given production process. A more controllable process is able to produce product wafers with higher yield and lower defect rates. Multiple factors can affect the etch rate profile, such as chamber pressure, temperature, power, and gas flow rates \cite{wu2010high, rickard2001characterization}. Thus, the optimization of etch rates is a complex process, often requiring extensive experimentation and metrology to draw correlations between etch rate profiles and tool output data. Multiple iterations are often required but experiments, materials, and metrology quickly become time-consuming and expensive.

Machine learning-based tools and techniques have become highly desirable across many industries as a means to improve efficiency by reducing labor, time, and cost demands. Machine learning (ML) can enable leaner, more resource-efficient semiconductor fabrication by improving the efficiency of fab processes, particularly complex and resource-intensive tasks such as etch rate optimization. This is especially true in environments where large volumes of process data are readily available for training and inference. In semiconductor manufacturing, ML techniques have been used to reduce defectivity \cite{cheng2021machine, kim2023advances, batool2021systematic}, in process management \cite{park2019reinforcement, li2012adaptive}, and in virtual metrology \cite{lenz2013data, hsieh2021convolutional}.

Data-driven modeling of physical processes provides an attractive alternative to conventional methods in other fields such as aerodynamics \cite{saetta2023abbottae, saetta2024uncertainty} and combustion \cite{chung2024ensemble}. In the present work, we describe a data-driven model that is developed to predict the second-order effects of a physical process; i.e. etch rates on a reactive ion etching tool. An autoencoder is used to encode the features of the etch rate profiles of different process conditions in the tool into a lower-dimensional representation, which is then used as a generative model by mapping the sensor input data into the lower-dimensional space. The error of the data-driven model is calculated both in the interpolation and extrapolation regime.

\section{Methodology}

\subsection{Dataset}

\begin{figure}

        \centering
        \includegraphics[width=0.9\linewidth,  trim = {0in 0in 0in 0in}, clip]{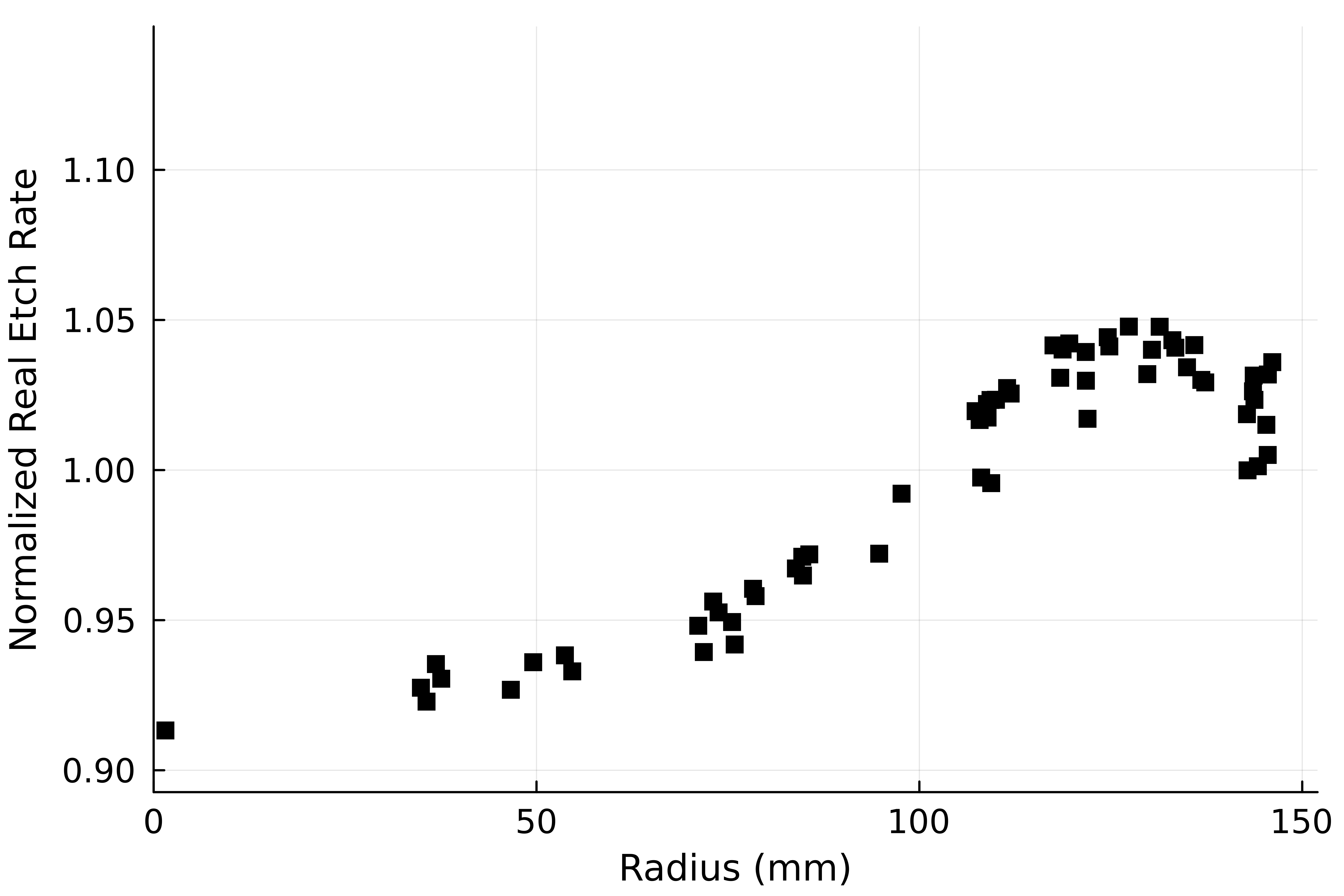}
    
    \caption{The measured etch rate against radius, for the case where [$\Delta \overset{\cdot}{m_1}$, $\Delta \overset{\cdot}{m_2}$, $\Delta P$] = [$0$, $0$, $0$].}
    \label{fig:interpolated_points}
\end{figure}

   $M = 17$ wafers processed on a single tool are considered for the dataset, where each datapoint is the etch rate measured on a single wafer at $N$ different locations [$X, Y$] on the surface, given by $W^{r,o}$. A plot of the etch rates against radius for the $N$ measurement points is depicted in \ref{fig:interpolated_points} for a sample wafer. Three sensor parameters of interest are considered, with $(\overset{\cdot}{m_1}, \overset{\cdot}{m_2})$ being the gas flow rate of the etching gas and the ashing gas into the chamber respectively and $P$ being the power delivered by the electromagnetic field. The process recipes included in the training dataset are shown in table \ref{tab:gas_flows}.

  \begin{table}[]
    \centering
    
    \begin{tabular}{|c|c|c|c|}
    \hline
         Sn. No & $\Delta \overset{\cdot}{m_1}$ $(\%)$ & $\Delta \overset{\cdot}{m_2}$ $(\%)$& $\Delta P$ $(\%)$\\
         \hline
         
         1& 0 & 0 & 0 \\
         \hline
         2& 0 & 10 & 0 \\
         3& 0 & 5 & 0 \\
         4& 0 & -5 & 0 \\
         5& 0 & -10 & 0 \\
         \hline
         6& 10 & 0 & 0 \\
         7& 5 & 0 & 0 \\
         8& -5 & 0 & 0 \\
         9& -10 & 0 & 0 \\
         \hline
         10& 10 & 10 & 0 \\
         11& 5 & 5 & 0 \\
         12& -10 & -10 & 0 \\
         13& -5 & -5 & 0 \\
         14& 5 & -5 & 0 \\
         15& -5 & 5 & 0 \\
         \hline
         16& 0 & 0 & -5 \\
         17& 0 & 0 & -10 \\
         
         \hline
    \end{tabular}
    \vspace{10pt}
    \caption{The combinations of change in gas flows from the original recipe ($\Delta\overset{\cdot}{m_1}$, $\Delta\overset{\cdot}{m_2}$) and power $\Delta P$ for the training wafers.}
    \label{tab:gas_flows}
\end{table}

\begin{figure}

        \centering
        \includegraphics[width=0.9\linewidth,  trim = {5in 0in 0in 0in}, clip]{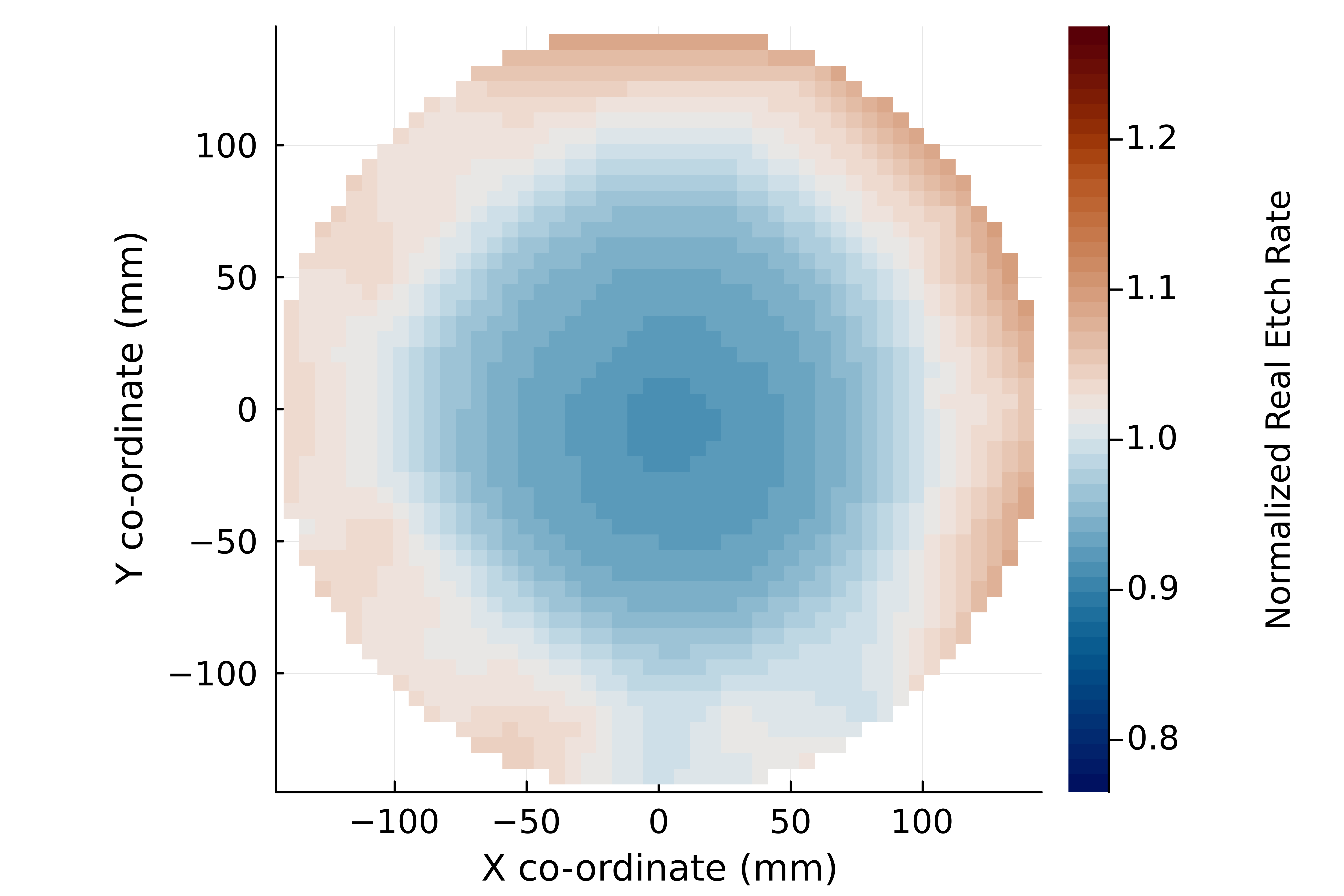}
    
    \caption{The normalized etch rate heatmap of the interpolated wafer, for the case where [$\Delta \overset{\cdot}{m_1}$, $\Delta \overset{\cdot}{m_2}$, $\Delta P$] = [$0$, $0$, $0$].}
    \label{fig:interpolated_wafer}
\end{figure}

 As the data is two-dimensional in nature, the $N$ points are used to generate a scalar field of etch rates on a grid of $n \times n$ points, with the scalar field being denoted as $W^{g,o}$. Multiquadric radial basis functions are used to generate this scalar field, which take the form:

\begin{equation}
    \varphi(r) = \sqrt{1 + (\epsilon r)^2}
\end{equation}

where $r$ is $||x - x_k|| \text{ } \forall  \text{ } k \in 1\dots N$, and $x$ is the position vector of the interpolation point with respect to a global origin. The cartesian grid of points that form the interpolants have the extents $[-R, R]$ where R is the radius of the wafer in each dimension, with a spacing of $R/n$. A heatmap of the scalar field is shown in Fig. \ref{fig:interpolated_wafer}. The input scalar field is then scaled to have a value between $0$ and $1$ by dividing the etch rate field by a scalar value that is equal to the maximum measured etch rate in the dataset. The interpolated scalar field is then used to train the autoencoder, with the autoencoder generating a new scalar field $W^{g,u}$. 

The input data also consists of the measured values from the tool's sensors for $\overset{\cdot}{m_1}$, $\overset{\cdot}{m_2}$ and $P$. We define $\Delta \overset{\cdot}{m_1}$, $\Delta \overset{\cdot}{m_2}$ and $\Delta P$ as the percent deviation from the POR recipe, for ease of representation of the cases. The sensor values are used instead of the recipe values under the assumption that the noise from the sensors is sufficiently minimal to not affect the model's predictions. The measured values are collected over the duration of the primary etch step during the RIE process. The mean value of the data over the etch step is then used to generate a $s \times 1$ vector for every wafer, where $s = 3$ is the number of input sensor parameters.

\subsection{Autoencoder}

\begin{figure}
    \centering
    \includegraphics[width=0.7\linewidth, trim = {1.5in 1.9in 1.8in 1in}, clip]{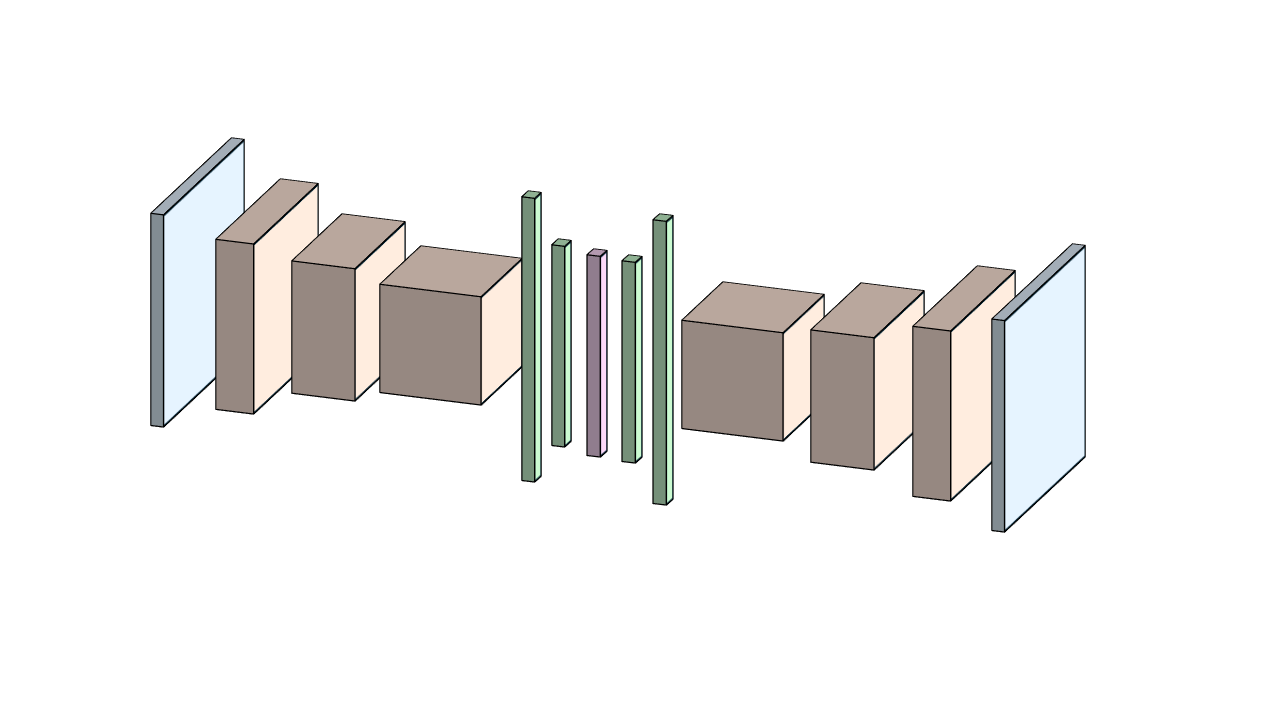}
    \caption{The convolutional autoencoder's architecture with the input and output heatmaps in blue, convolutional layers in beige, fully connected layers in green and the latent space in lilac.}
    \label{fig:autoencoder-architecture}
\end{figure}

Autoencoders are a class of ML algorithms that are used to efficiently encode features of unlabeled higher-dimensional data into a lower-order representation. Autoencoders consist of two models - an encoder which is used to compress the input data into the lower-dimensional vector that is called a latent space, and the decoder which is used to regenerate the input data. The construction of an autoencoder can be mathematically summarized as:

\begin{equation}
    \overline{l} = E(X)
\end{equation}

and

\begin{equation}
    X' = D(\overline{l})
\end{equation}

where $X$ is the input data, $X'$ is the generated data, $\overline{l}$ is the latent space vector, $E(\cdot)$ is the encoder and $D(\cdot)$ is the decoder. We also note that 

\begin{equation}
    X, X' \in \mathbb{R}^d
\end{equation}

and

\begin{equation}
    \overline{l} \in \mathbb{R}^p
\end{equation}

where $d$ is the dimension of the input data and $p$ the dimension of the latent space. The autoencoder architecture is described in fig. \ref{fig:autoencoder-architecture}. Here, we use an autoencoder to encode the features of the etch rate profiles on the wafer into the latent space vector. We describe the model as a convolutional autoencoder, as multiple convolutional layers are used in both the encoder and decoder models to extract two-dimensional patterns in the etch rate field. Three convolutional layers are used, with the filter sizes for each of the convolutional layers being $5$, $3$ and $2$ with the number of channels as $50$, $100$ and $200$ respectively. These layers feed into two fully connected layers, with the first layer having the same dimensions as the number of weights in the last convolution layer, and the second fully connected layer having the same dimensions as the latent space. The tanh activation function is used throughout the model. The parameters of the model are designated as $w_e$, with the total number of weights being $n_e$.

The loss function for training the autoencoder is defined as

\begin{equation}
    \mathcal{L}(W^{g,o}, W^{g,u}) =   l_1(W^{g,o}, W^{g,u}) +  l_2(W^{g,o}, W^{g,u}) + l_3(w_e)
\end{equation}

where 

\begin{equation}
    l_1(W^{g,o}, W^{g,u}) =  \frac{\alpha_1}{B}\sum_{i = 1}^B \sqrt{(W^{g,o}_i - W^{g,u}_i)^2}  
\end{equation}

\begin{equation}
    l_2(W^{g,o}, W^{g,u}) = \frac{\alpha_2}{B}\sum_{i = 1}^B max(|W_i^{g,o} - W_i^{g,u}|)
\end{equation}

\begin{equation}
    l_3(w_e) = \frac{\alpha_3}{B}\sum_{i = 1}^B  \frac{1}{n_e} \sqrt{\sum_{j = 1}^{n_e} w_{e,(i,j)}^2}
\end{equation}

where $B$ is the batch size. While $l_1$ is the commonly used root mean square error, $l_2$ is the absolute maximum of the difference between the model-generated wafer maps and the interpolated wafer map. A common issue in the training of autoencoders is that the model learns to generate the average of the batch to minimize $l_1$. The addition of $l_2$ in the loss function avoids this problem, and the inclusion of $l_3$ prevents overfitting. We define $\alpha_1$ and $\alpha_2$ such that

\begin{equation}
    \alpha_1 + \alpha_2 = 1
\end{equation}

with the value of $\alpha_2$ being discretely selected as $[0.05, 0.005, 0]$ as training progresses. $\alpha_3$ is initially set as $\alpha_3 = 6.7 \times 10^{-3}$, and set to $\alpha_3 = 0$ as the loss approaches convergence. For the gradient descent, a staged approach is used with the ADAM optimizer.


Since convolutions on a cartesian grid are more natural than convolutions on other grids, the interpolation is carried out to generate the scalar field at cartesian coordinates. Since spurious data is generated outside the radial wafer bounds a final post-processing step is applied where all the data outside the wafer radius is discarded after generation and loss calculation. We also normalize the etch rate from all predictions with the mean etch rate of the wafer that is the point of reference.

\subsection{Mapping}

In order to map the sensor variables to an output etch rate wafermap, we employ an algorithm where the encoder is used to generate the latent space for every known wafer, and each of the latent space variables $l_h$ are mapped onto the sensor space $\mathbb{R}_s$ with interpolation using multiquadric radial basis functions as

\begin{equation}
    l_h = \mathbb{F}_h(\overset{\cdot}{m_1}, \overset{\cdot}{m_2}, P) \text{  }\forall\text{  } h \text{  }\in \text{  }1\dots p
\end{equation}

 This mapping results in the creation of a scalar field in three dimensions for each of the latent space variables, much akin to the scalar field in two dimensions that is generated for the etch rate heatmaps, with $p$ scalar fields being generated for the entire latent space vector. Using these maps, we then determine the new latent space variables for any value of the sensor parameters through interpolation and extrapolation. Post this, the latent space vector is passed through the decoder to generate synthetic wafermaps, denoted by $W^{g,s}$ at the interpolation points. The final surrogate model generating the etch rate profiles from the sensor data is defined by

 \begin{equation}
   W_t^{g,s}  = D([\mathbb{F}_{h,t}(\overset{\cdot}{m_{1,t}}, \overset{\cdot}{m_{2,t}}, P_t)]) \text{  }\forall\text{  } h \text{  }\in \text{  }1\dots p
\end{equation}

evaluated at values outside the training set $\overset{\cdot}{m_{1,t}}$, $\overset{\cdot}{m_{2,t}}$ and $P_t$.

\begin{table}[]
    \centering
    
    \begin{tabular}{|c|c|c|c|}
    \hline
         Sn. No & $\Delta \overset{\cdot}{m_1}$ $(\%)$ & $\Delta \overset{\cdot}{m_2}$ $(\%)$& $\Delta P$ $(\%)$\\
         \hline
         18& 12 & 0 & 0 \\

         19& 7 & -3 & 0 \\
                  
         20& -2 & 7 & 0 \\

         21& 4 & 0 & 0 \\

         22& 0 & 0 & -6 \\

         \hline
    \end{tabular}
    \vspace{10pt}
    \caption{The combinations of change in gas flows from the original recipe ($\Delta\overset{\cdot}{m_1}$, $\Delta\overset{\cdot}{m_2}$) and power $\Delta P$ for the testing wafers.}
    \label{tab:gas_flows_test}
\end{table}

\subsection{Wafer statistics calculation}

As the value of the etch rate field is exactly defined at the measurement points but not at the interpolated heatmap grid, we perform a final interpolation step to calculate the mean for the synthetic wafer maps. We take the generated wafermaps of $n \times n$ datapoints and once again apply the multiquadric radial basis function to interpolate to the $X$ and $Y$ coordinates of the original N datapoints, giving us a new scalar field $W^{r,u}$ for the autoencoder model and $W^{r,s}$ for the sensor mapping model.  The mean of the etch rate on the wafer is then calculated as

\begin{equation}
    \mu(W^{r,u}) = \frac{1}{N}\sum_{i = 1}^N W^{r,u}_i
\end{equation}

and

\begin{equation}
    \mu(W^{r,s}) = \frac{1}{N}\sum_{i = 1}^N W^{r,s}_i
\end{equation}

respectively. This method of calculating the etch rate mean  allows for exact comparison with etch rate statistics derived from metrology.

 \begin{figure}
     \centering
     \includegraphics[width=0.9\linewidth]{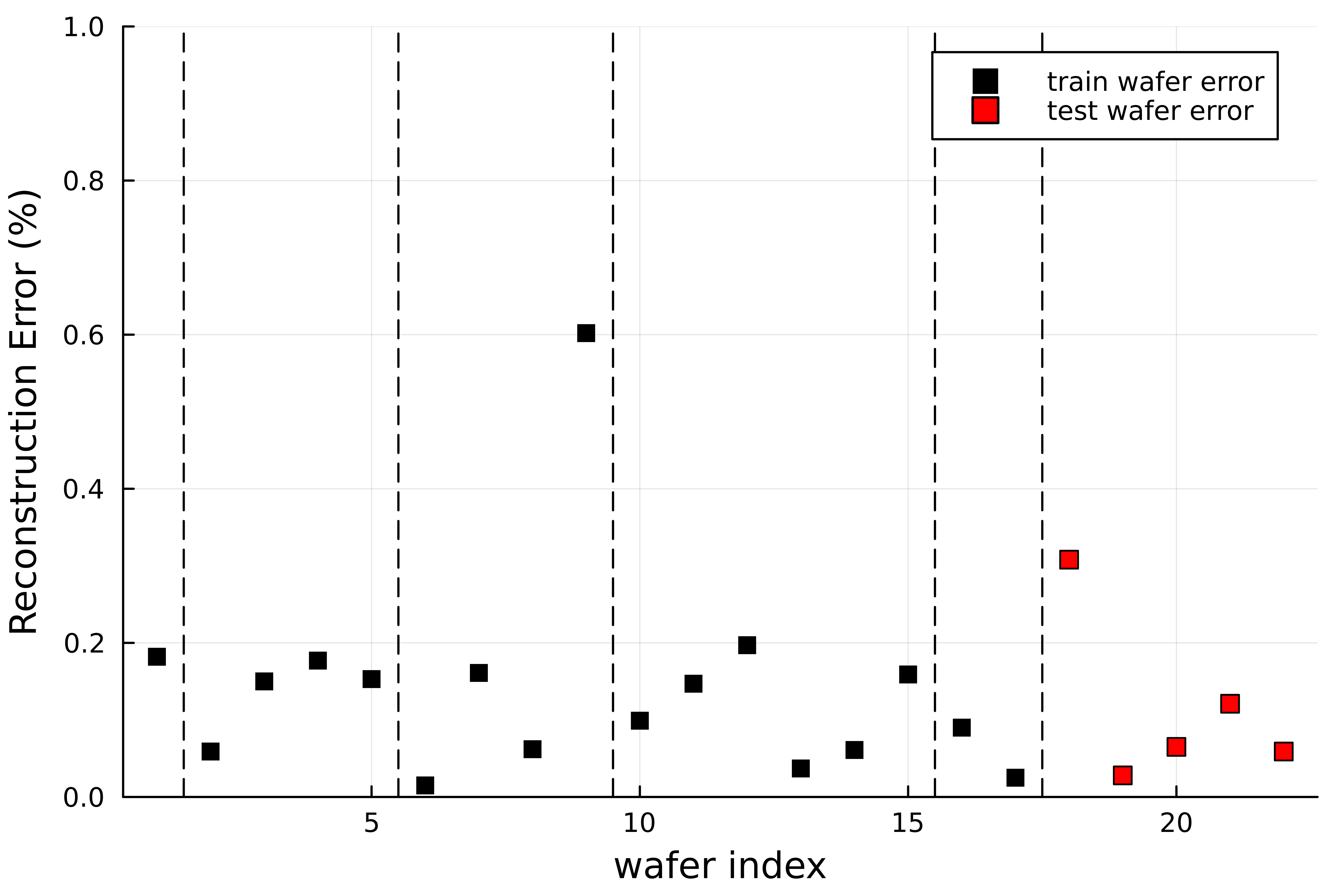}
     \caption{The mean reconstruction error $\mathcal{E}_{r,\mu}$ for the train and test wafers, evaluated at the measurement points.}
     \label{fig:reconstruction_error_all_wafers}
 \end{figure}

\section{Results}

We define four measures to evaluate the model's accuracy:  
the local reconstruction error $\mathcal{E}_{r,i}$, the error in the mean from reconstruction $\mathcal{E}_{r,\mu}$, the local sensor-predicted error $\mathcal{E}_{s,i}$, and the sensor-predicted mean etch rate error $\mathcal{E}_{s,\mu}$. These measures are given by

\begin{equation}
    \mathcal{E}_{r,i} = \frac{|W^{g,u}_i - W^{g,o}_i|}  {W^{g,o}_i} 
\end{equation}

\begin{equation}
    \mathcal{E}_{r,\mu} = \frac{1}{N} \sum_{i=1}^N \frac{|W^{r,u}_i - W^{r,o}_i| }{W^{r,o}_i}
\end{equation}

\begin{equation}
    \mathcal{E}_{s,i} = \frac{|W^{g,s}_i - W^{g,o}_i| }{W^{g,o}_i}
\end{equation}

\begin{equation}
    \mathcal{E}_{s,\mu} = \frac{1}{N} \sum_{i=1}^N \frac{|W^{r,s}_i - W^{r,o}_i|}{W^{r,o}_i} 
\end{equation}

 Five new recipes are used to test the model, which are described in table \ref{tab:gas_flows_test}. The first of these recipes is extrapolatory in nature, i.e. the parameters for the recipe exist outside the training sensor space, with the rest of the recipes being interpolatory. As the surrogate model is expected to perform worse in extrapolation rather than intrapolation, these recipes are designed to test both cases.

\begin{figure*}

    \centering

    \begin{subfigure}[t!]{0.33\linewidth}
        \centering
        \includegraphics[width=\linewidth,  trim = {0in 0in 0in 0in}, clip]{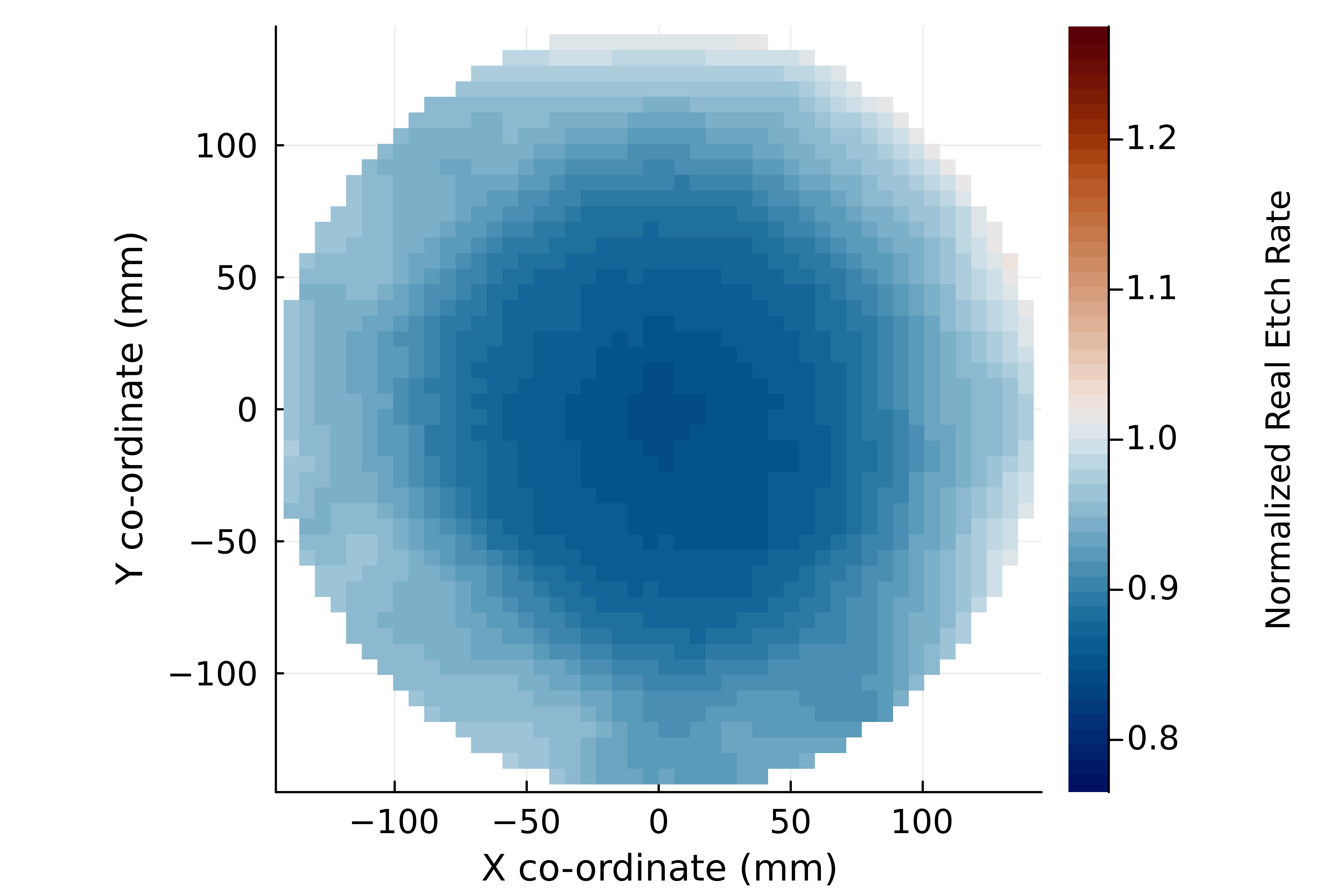}
    \caption{ }
    \end{subfigure}
~
    \begin{subfigure}[t!]{0.33\linewidth}
        \centering
        \includegraphics[width=\linewidth,  trim = {0in 0in 0in 0in}, clip]{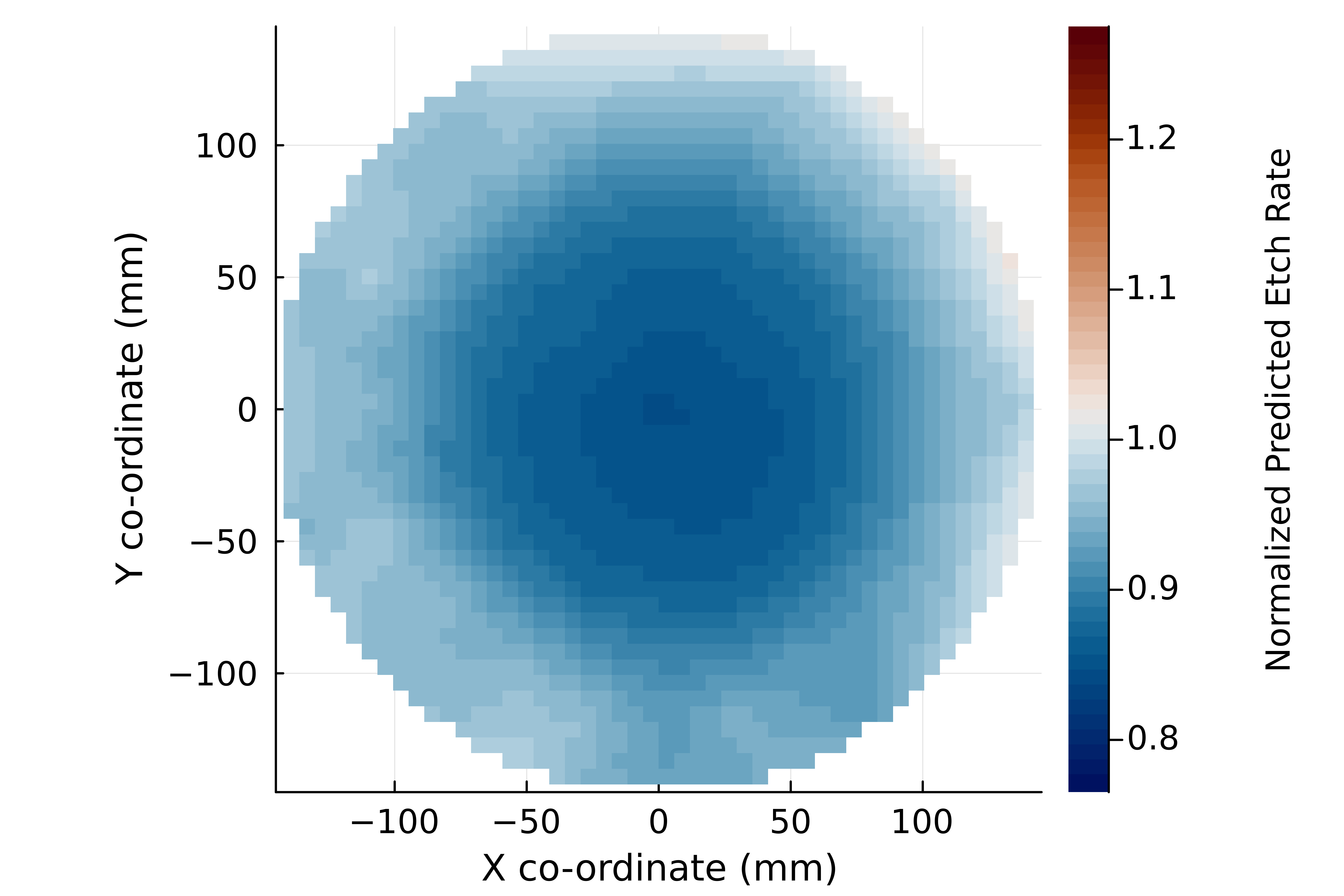}
    \caption{ }
    \end{subfigure}
~
    \begin{subfigure}[t!]{0.3\linewidth}
    \centering
    \includegraphics[width=\linewidth,  trim = {0in 0in 0in 0in}, clip]{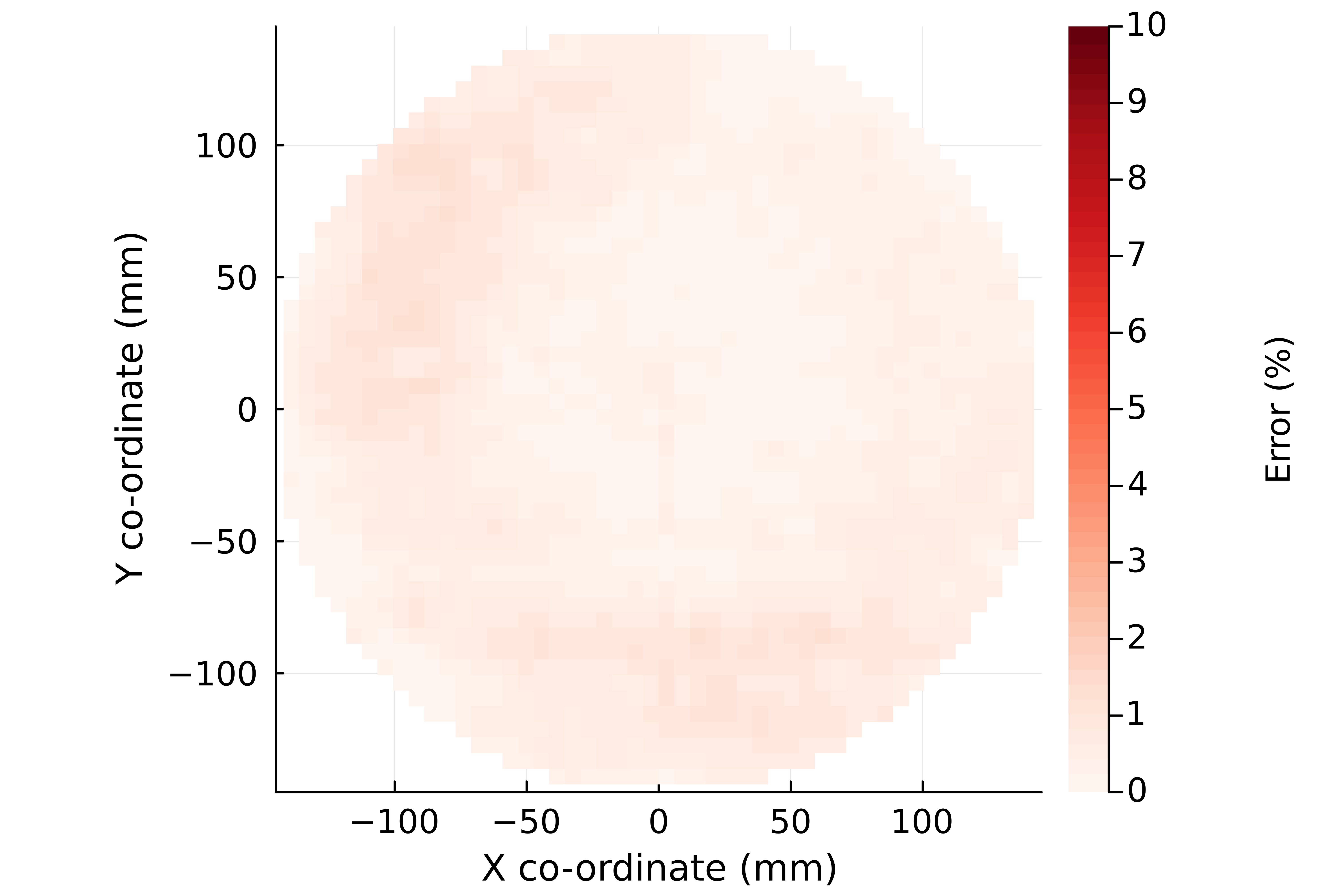}
    \caption{ }
    \end{subfigure}

        \centering

    \begin{subfigure}[t!]{0.33\linewidth}
        \centering
        \includegraphics[width=\linewidth,  trim = {0in 0in 0in 0in}, clip]{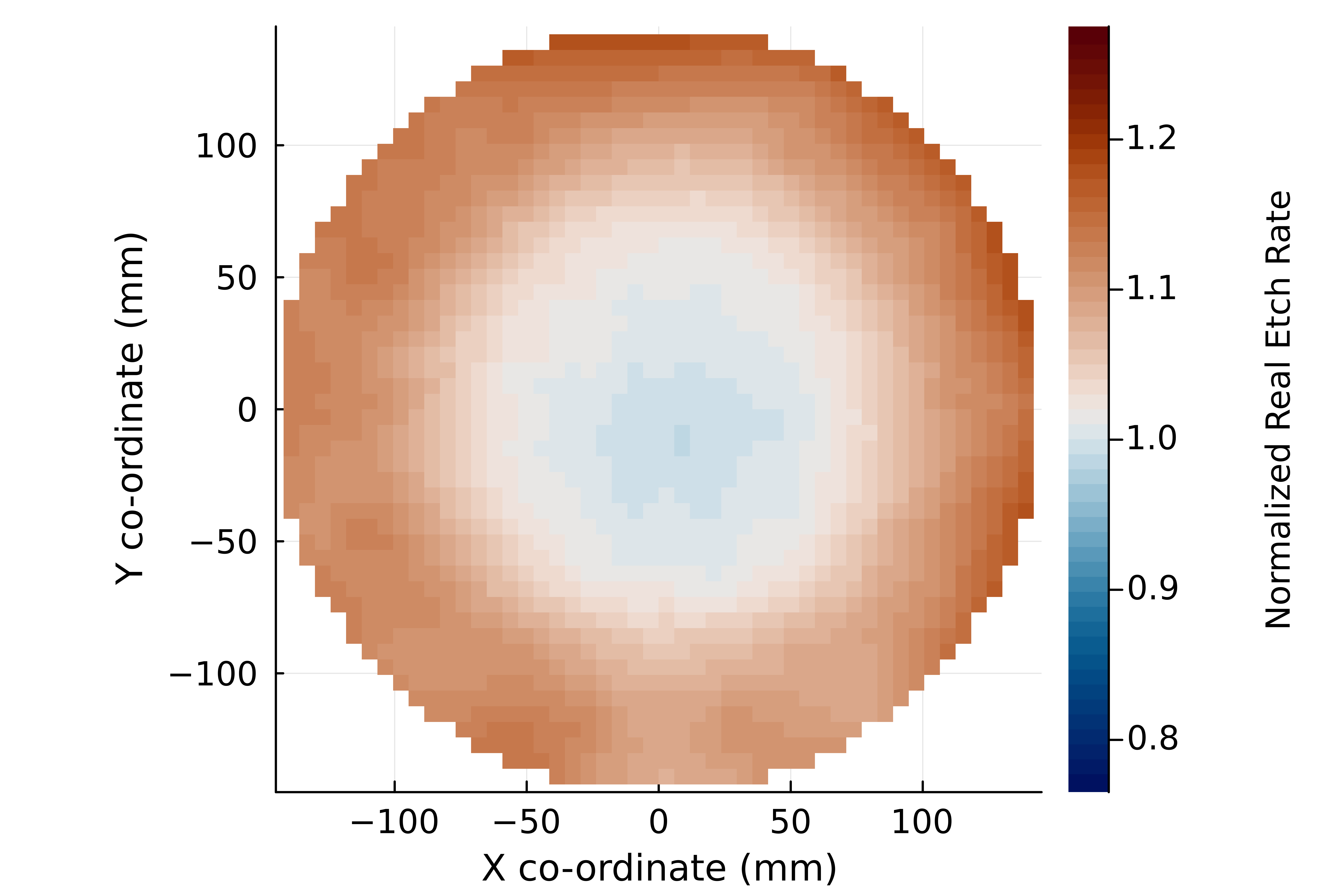}
    \caption{ }
    \end{subfigure}
~
    \begin{subfigure}[t!]{0.33\linewidth}
        \centering
        \includegraphics[width=\linewidth,  trim = {0in 0in 0in 0in}, clip]{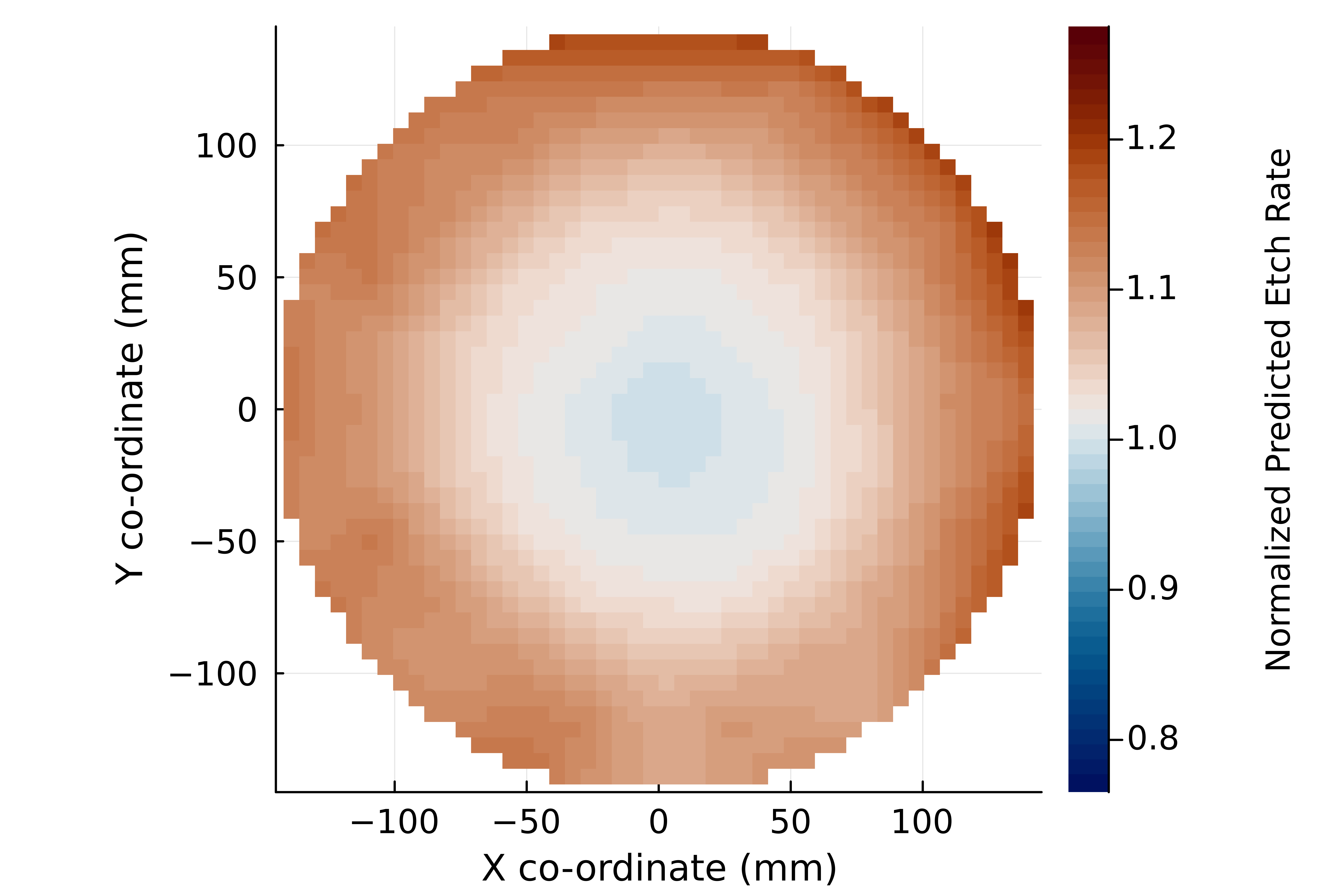}
    \caption{ }
    \end{subfigure}
~
    \begin{subfigure}[t!]{0.3\linewidth}
    \centering
    \includegraphics[width=\linewidth,  trim = {0in 0in 0in 0in}, clip]{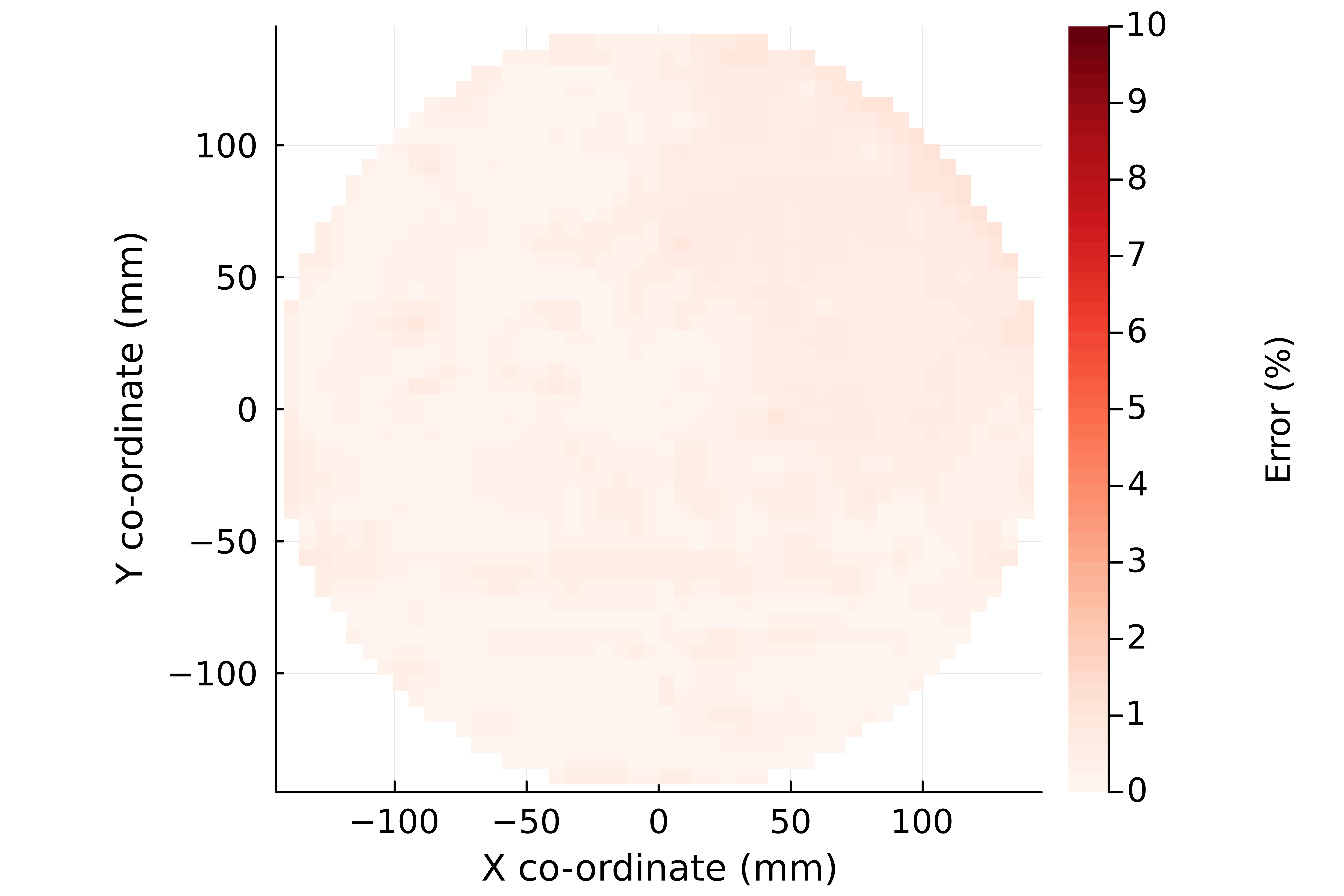}
    \caption{ }
    \end{subfigure}

    \caption{(a,b,c) The measured etch rate heatmap, the reconstructed etch rate heatmap and the local reconstruction error $\mathcal{E}_{r,i}$  for the case where [$\Delta \overset{\cdot}{m_1}$, $\Delta \overset{\cdot}{m_2}$, $\Delta P$] = [$-10$, $0$, $0$] (d,e,f) the same for the case where  [$\Delta \overset{\cdot}{m_1}$, $\Delta \overset{\cdot}{m_2}$, $\Delta P$] = [$12$, $0$, $0$]  }
    \label{fig:reconstruction_error_comparison}
\end{figure*}

 We first examine the reconstruction error for the dataset in fig. \ref{fig:reconstruction_error_all_wafers}. The first 17 datapoints in black correspond to the training set, and the last five datapoints in red to the test set. We note that the reconstruction error is below $0.2\%$ for almost all the datapoints in both the training as well as the testing set, with the cases where [$\Delta \overset{\cdot}{m_1}$, $\Delta \overset{\cdot}{m_2}$, $\Delta P$] = [$-10$, $0$, $0$] and [$\Delta \overset{\cdot}{m_1}$, $\Delta \overset{\cdot}{m_2}$, $\Delta P$] = [$12$, $0$, $0$] being the outliers. The wafermaps and reconstruction errors for these two cases are shown in fig. \ref{fig:reconstruction_error_comparison}.

\begin{figure}
    \centering
    \includegraphics[width=0.9\linewidth]{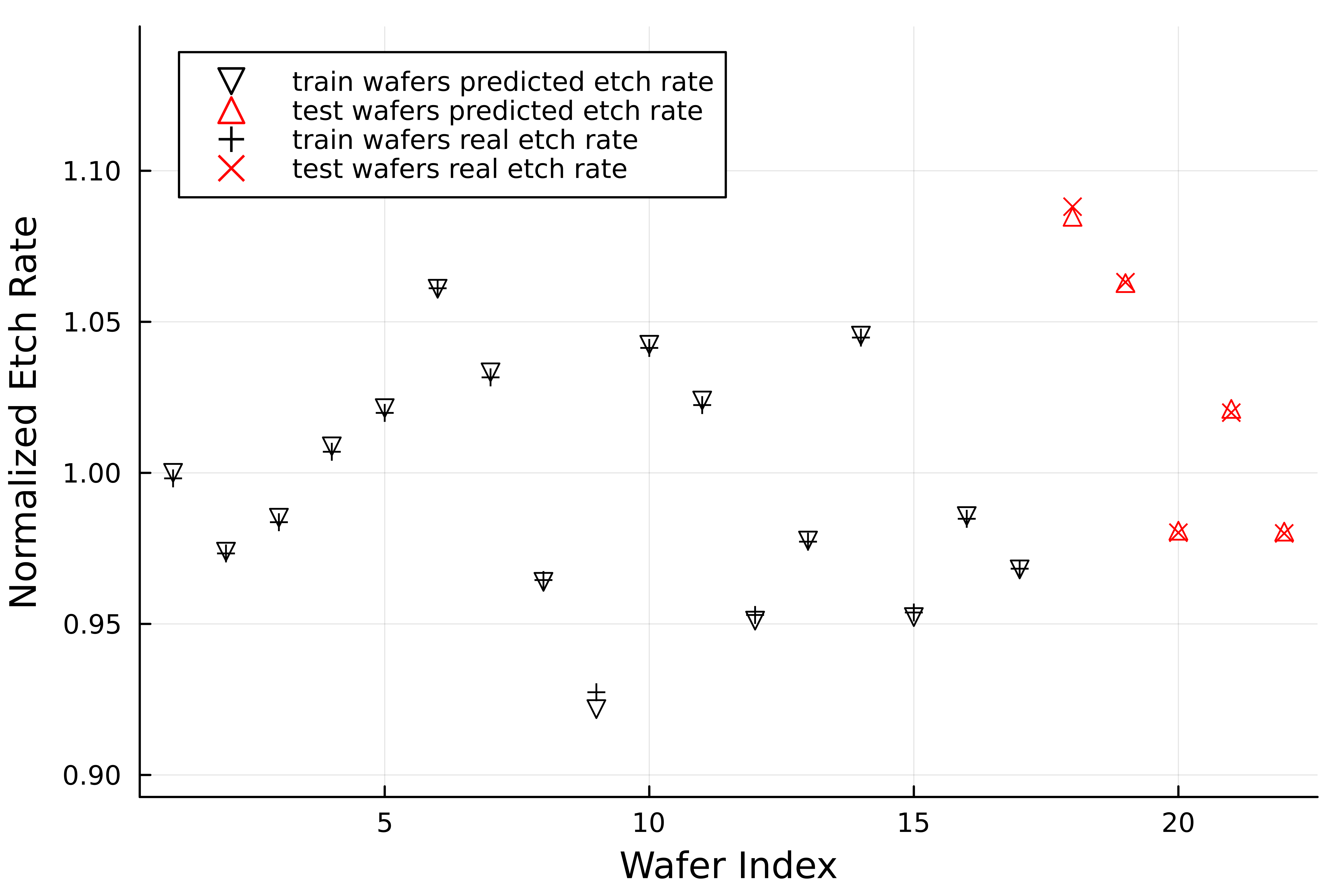}
    \caption{The sensor-predicted ($\mu(W^{r,s})$) and real mean etch rates ($\mu(W^{r,o})$) for the test wafers by the surrogate model}
    \label{fig:etch_rate_AE}
\end{figure}

 On closer inspection of the etch rate predictions and true etch rates from the autoencoder in fig. \ref{fig:etch_rate_AE}, we observe that the worst error is observed for the cases in the dataset with the highest and lowest mean etch rate. We also note that the etch rate prediction is far worse for the case where the datapoint is still in the training set and the etch rate is lower than the etch rate for all other training datapoints. 

\begin{figure*}

    \centering

    \begin{subfigure}[t!]{0.33\linewidth}
        \centering
        \includegraphics[width=\linewidth,  trim = {0in 0in 0in 0in}, clip]{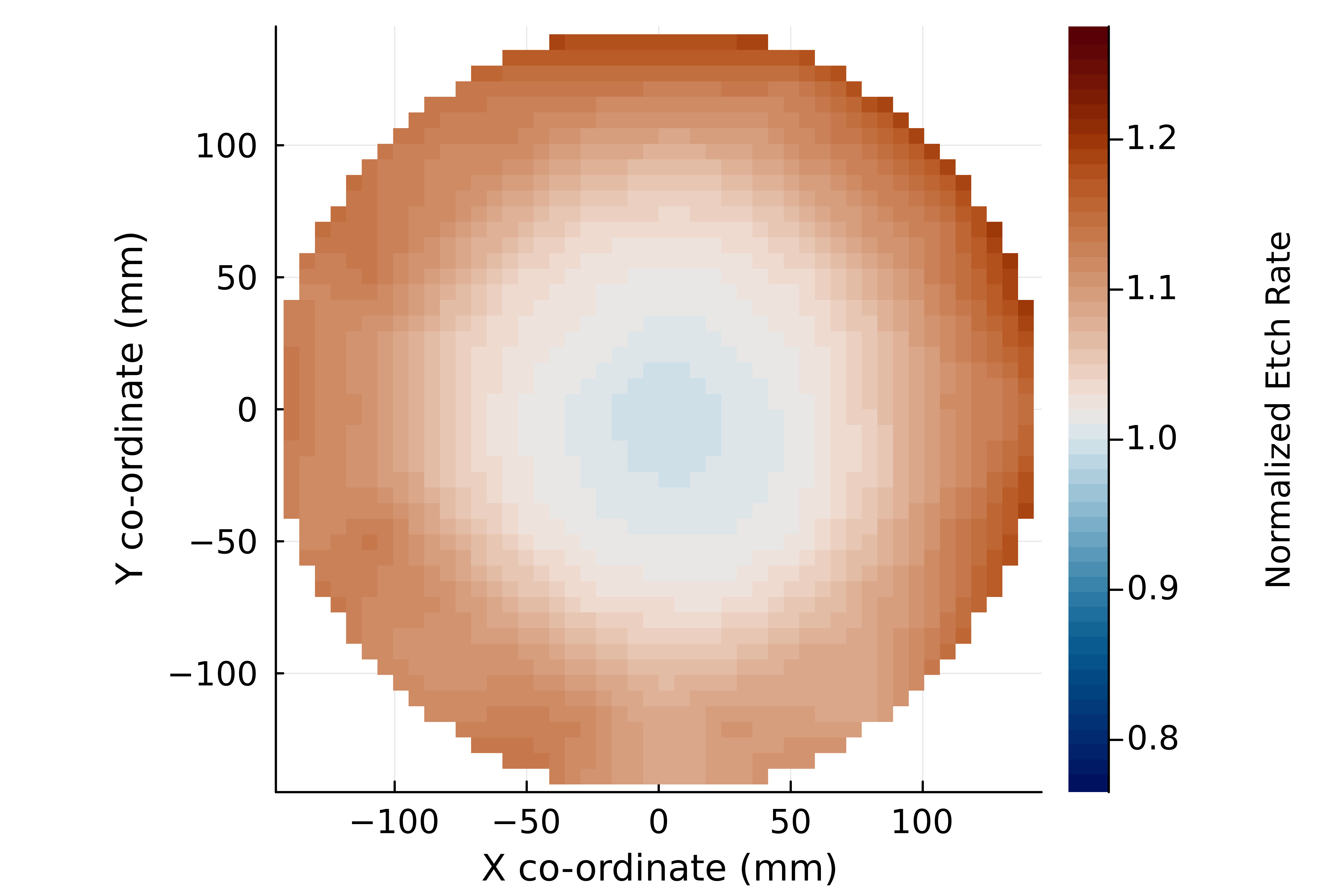}
    \caption{ }
    \end{subfigure}
~
    \begin{subfigure}[t!]{0.33\linewidth}
        \centering
        \includegraphics[width=\linewidth,  trim = {0in 0in 0in 0in}, clip]{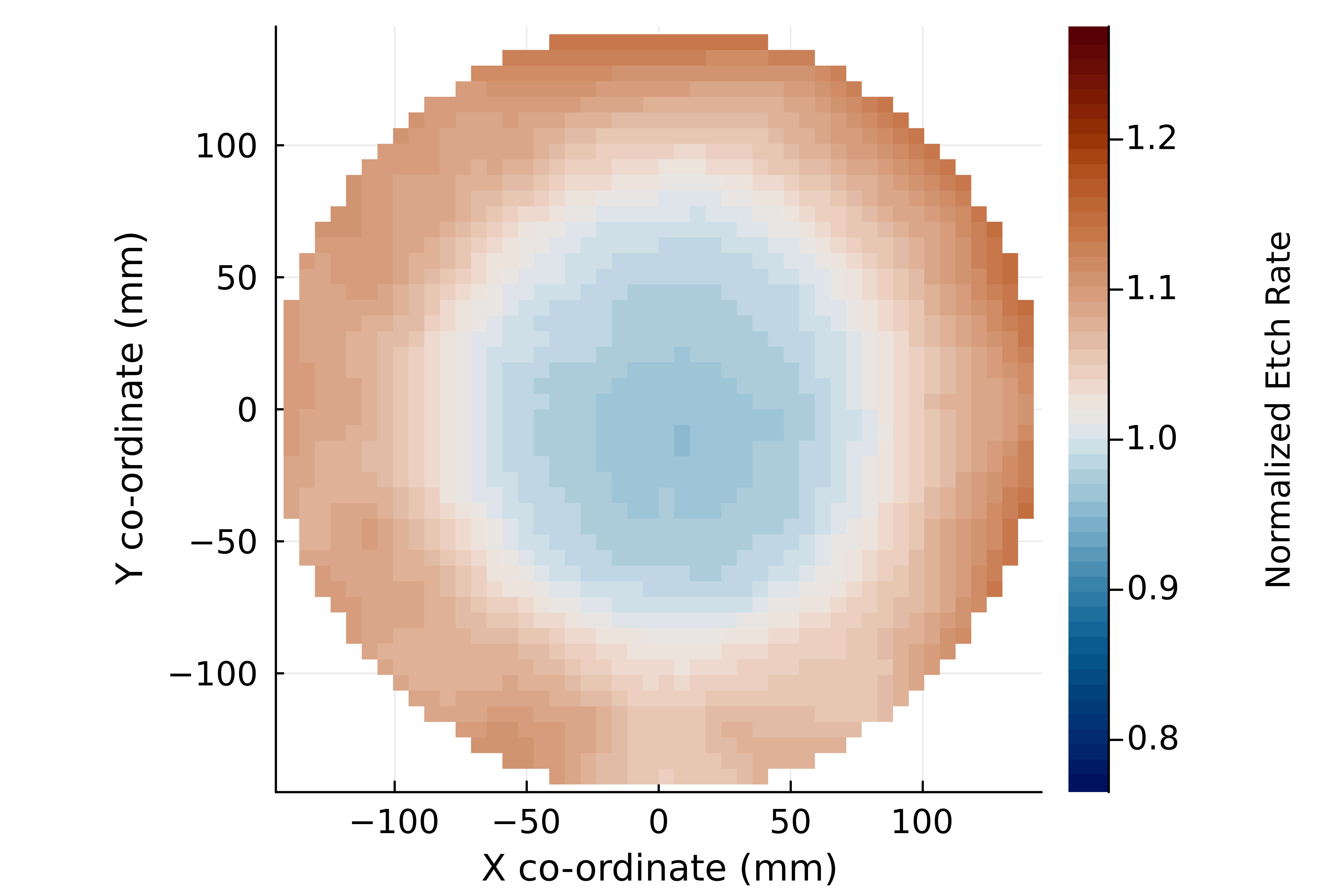}
    \caption{ }
    \end{subfigure}
~
    \begin{subfigure}[t!]{0.3\linewidth}
    \centering
    \includegraphics[width=\linewidth,  trim = {0in 0in 0in 0in}, clip]{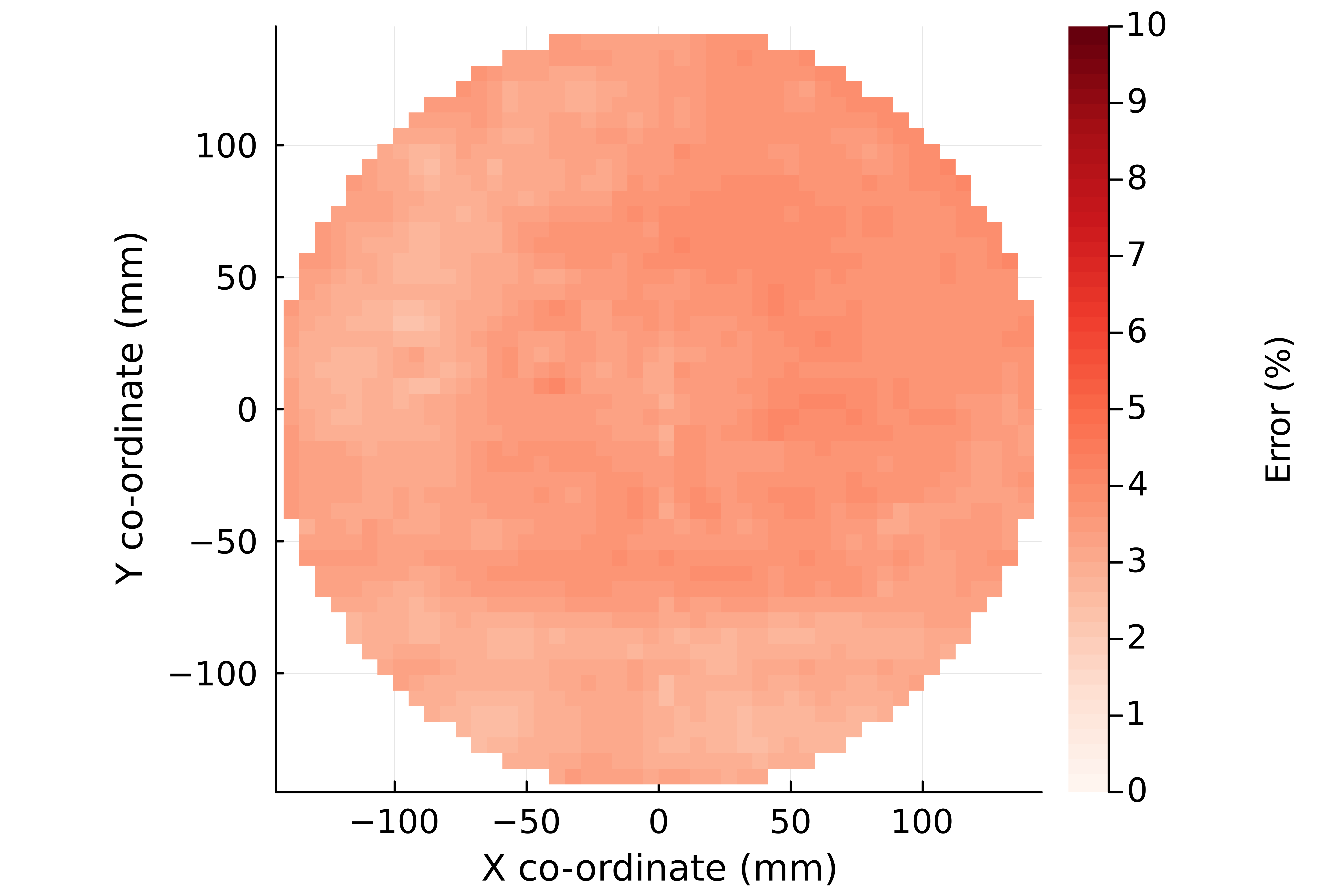}
    \caption{ }
    \end{subfigure}

    \centering

    \begin{subfigure}[t!]{0.33\linewidth}
        \centering
        \includegraphics[width=\linewidth,  trim = {0in 0in 0in 0in}, clip]{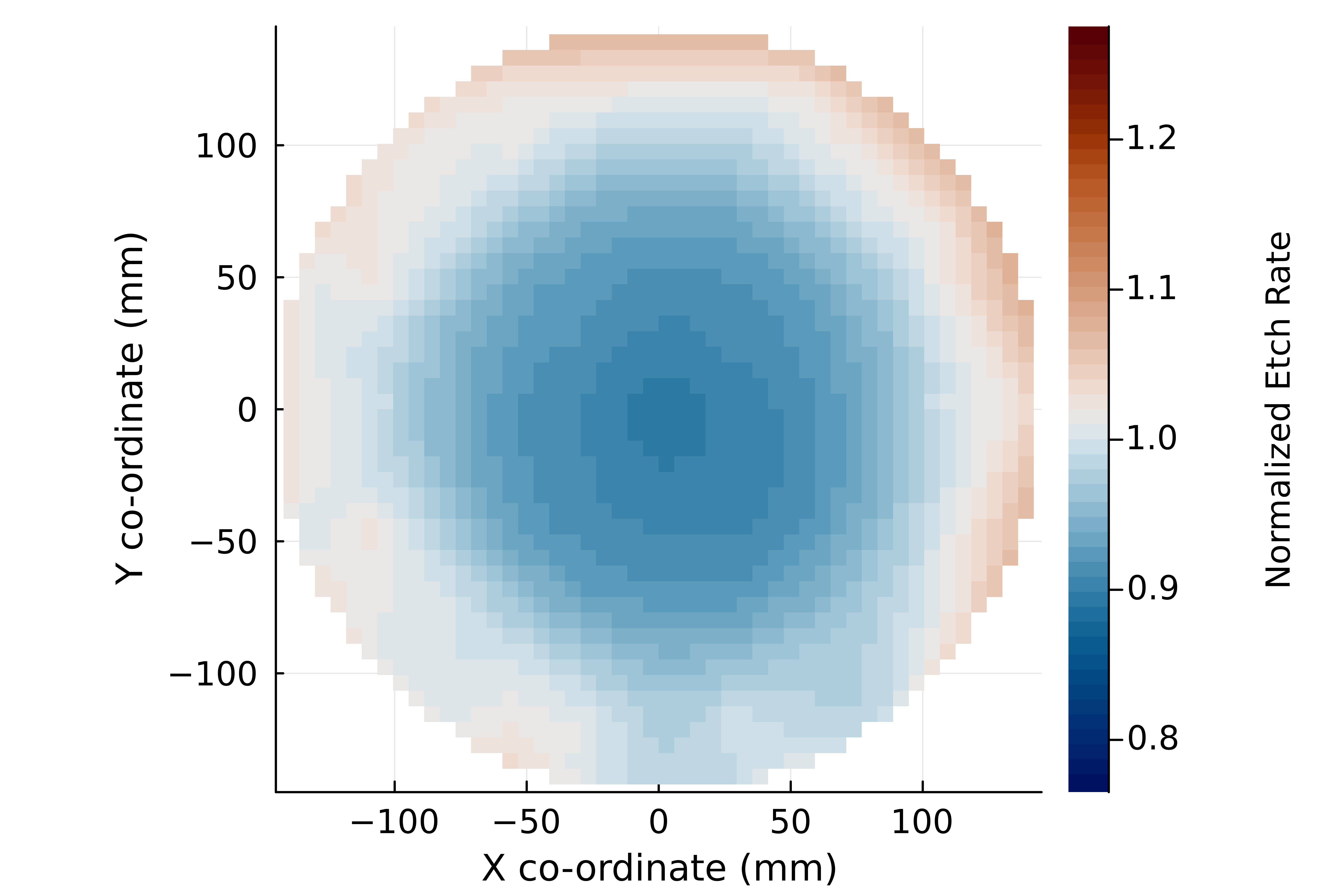}
    \caption{ }
    \end{subfigure}
~
    \begin{subfigure}[t!]{0.33\linewidth}
        \centering
        \includegraphics[width=\linewidth,  trim = {0in 0in 0in 0in}, clip]{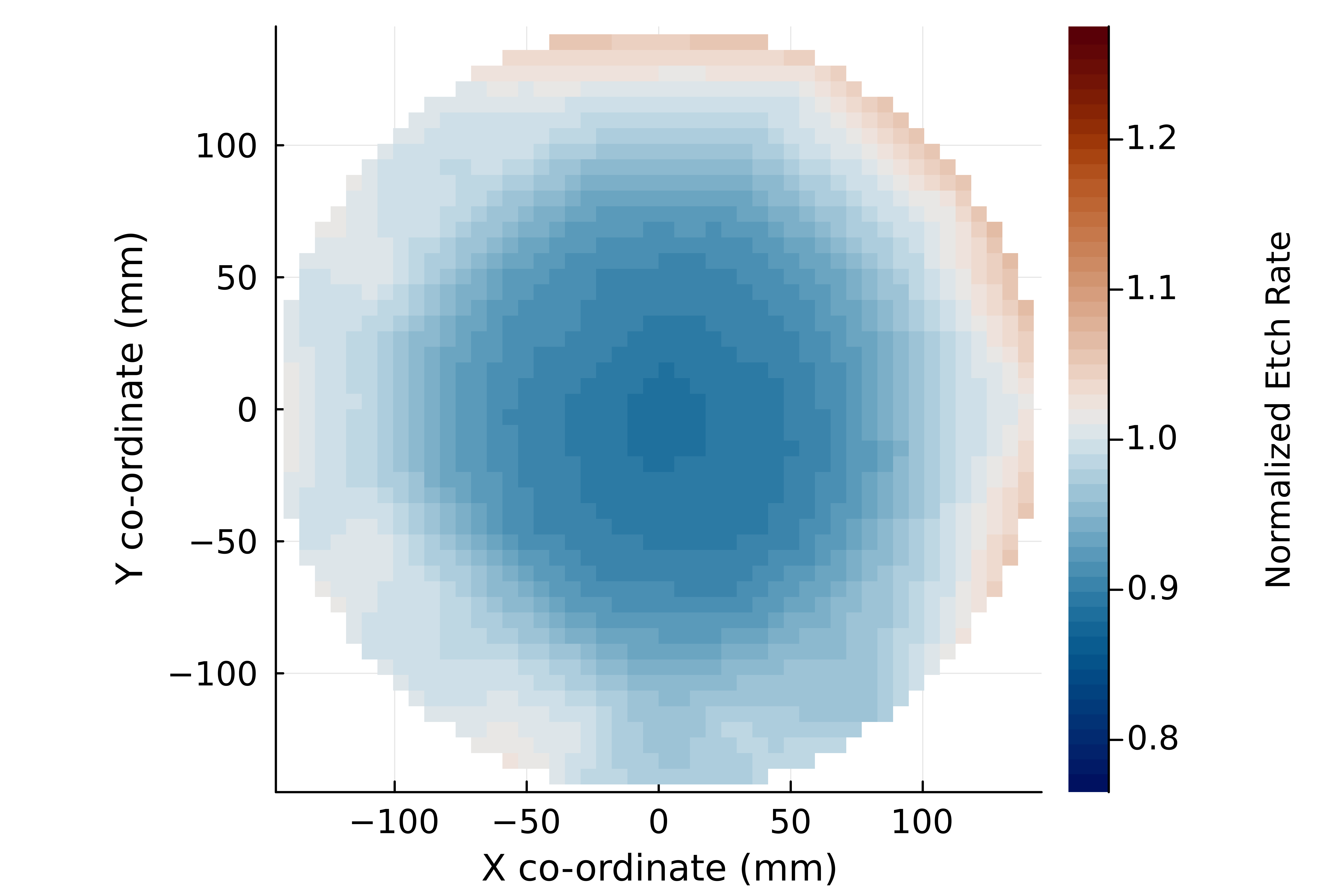}
    \caption{ }
    \end{subfigure}
~
    \begin{subfigure}[t!]{0.3\linewidth}
    \centering
    \includegraphics[width=\linewidth,  trim = {0in 0in 0in 0in}, clip]{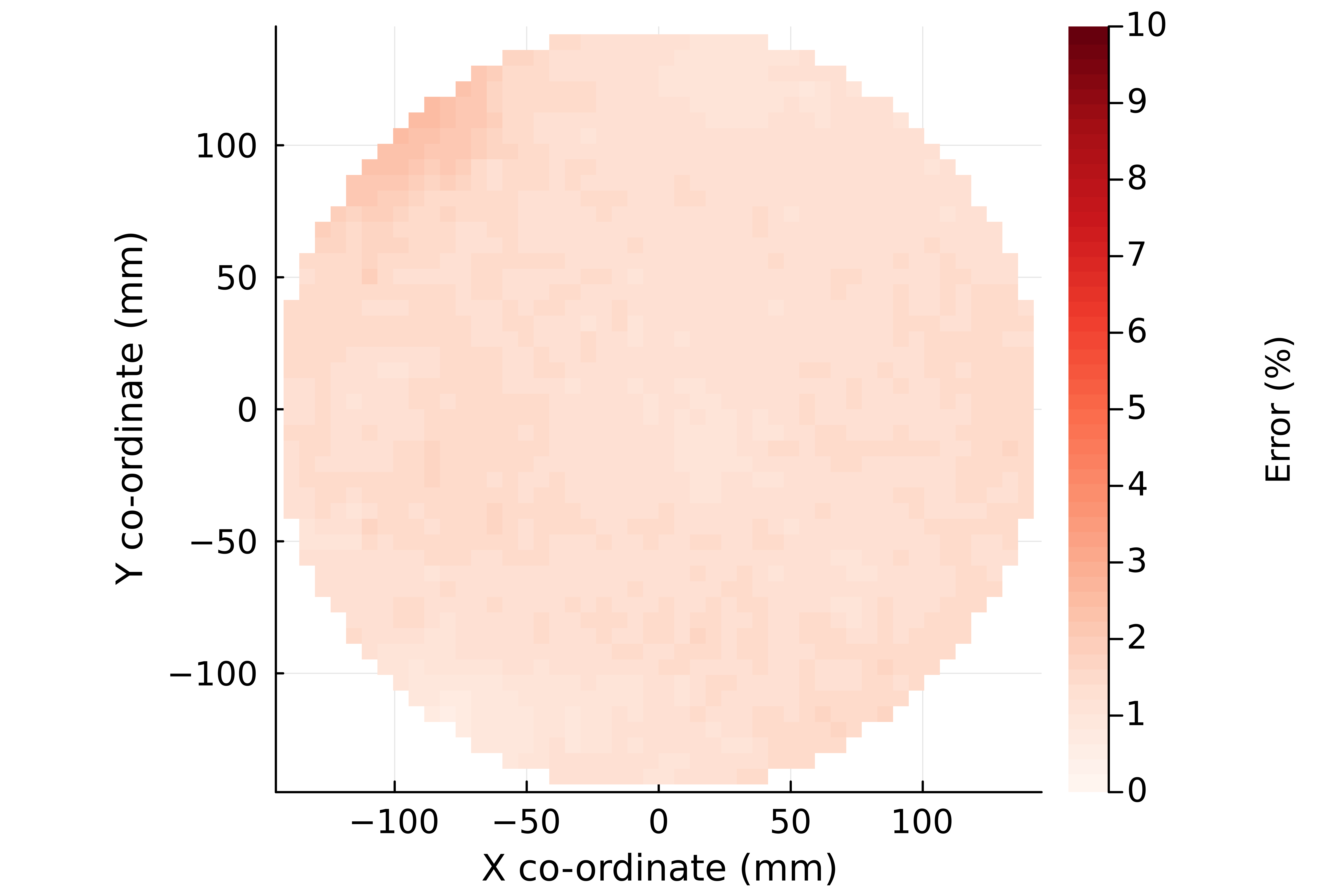}
    \caption{ }
    \end{subfigure}

    \centering

    \begin{subfigure}[t!]{0.33\linewidth}
        \centering
        \includegraphics[width=\linewidth,  trim = {0in 0in 0in 0in}, clip]{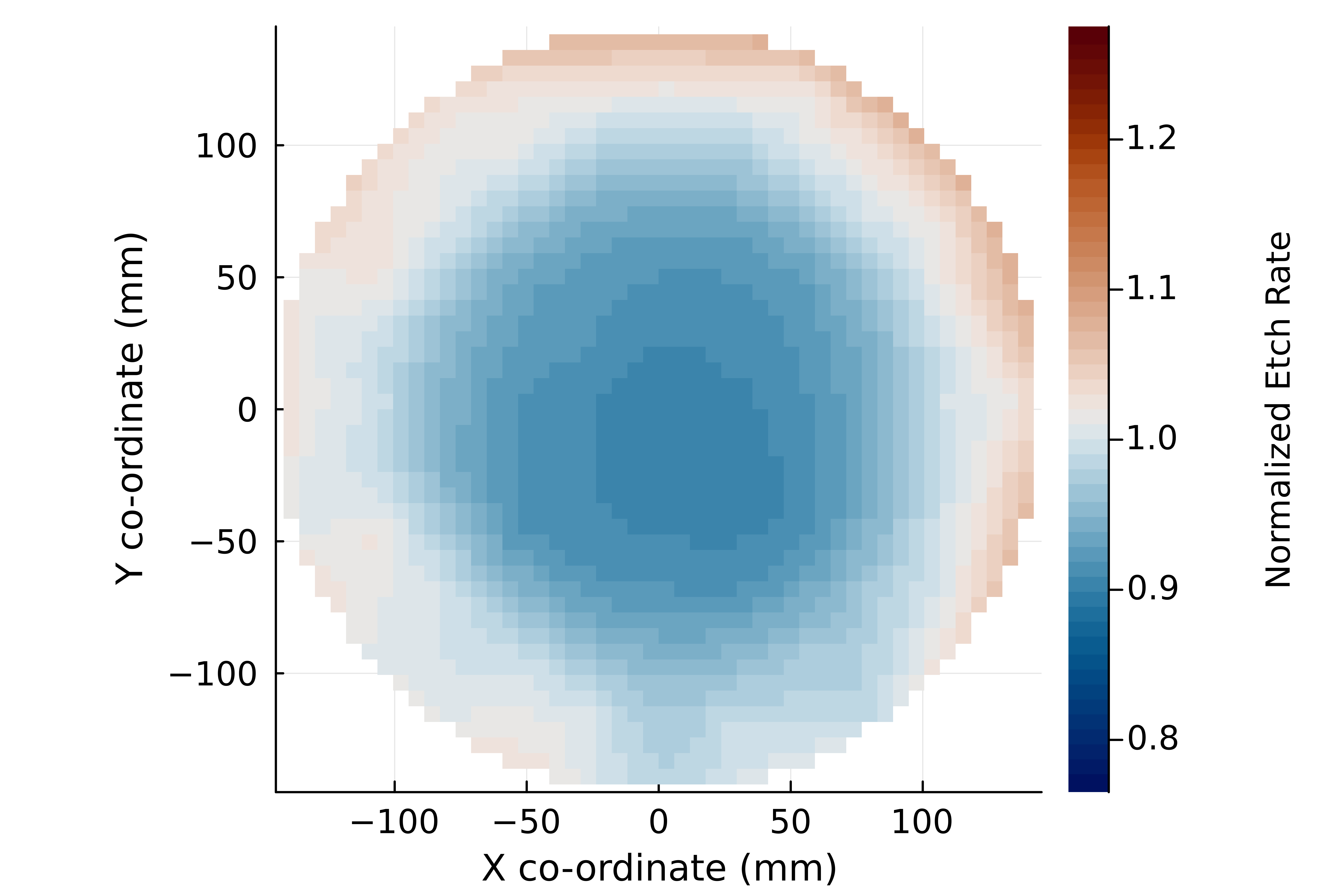}
    \caption{ }
    \end{subfigure}
~
    \begin{subfigure}[t!]{0.33\linewidth}
        \centering
        \includegraphics[width=\linewidth,  trim = {0in 0in 0in 0in}, clip]{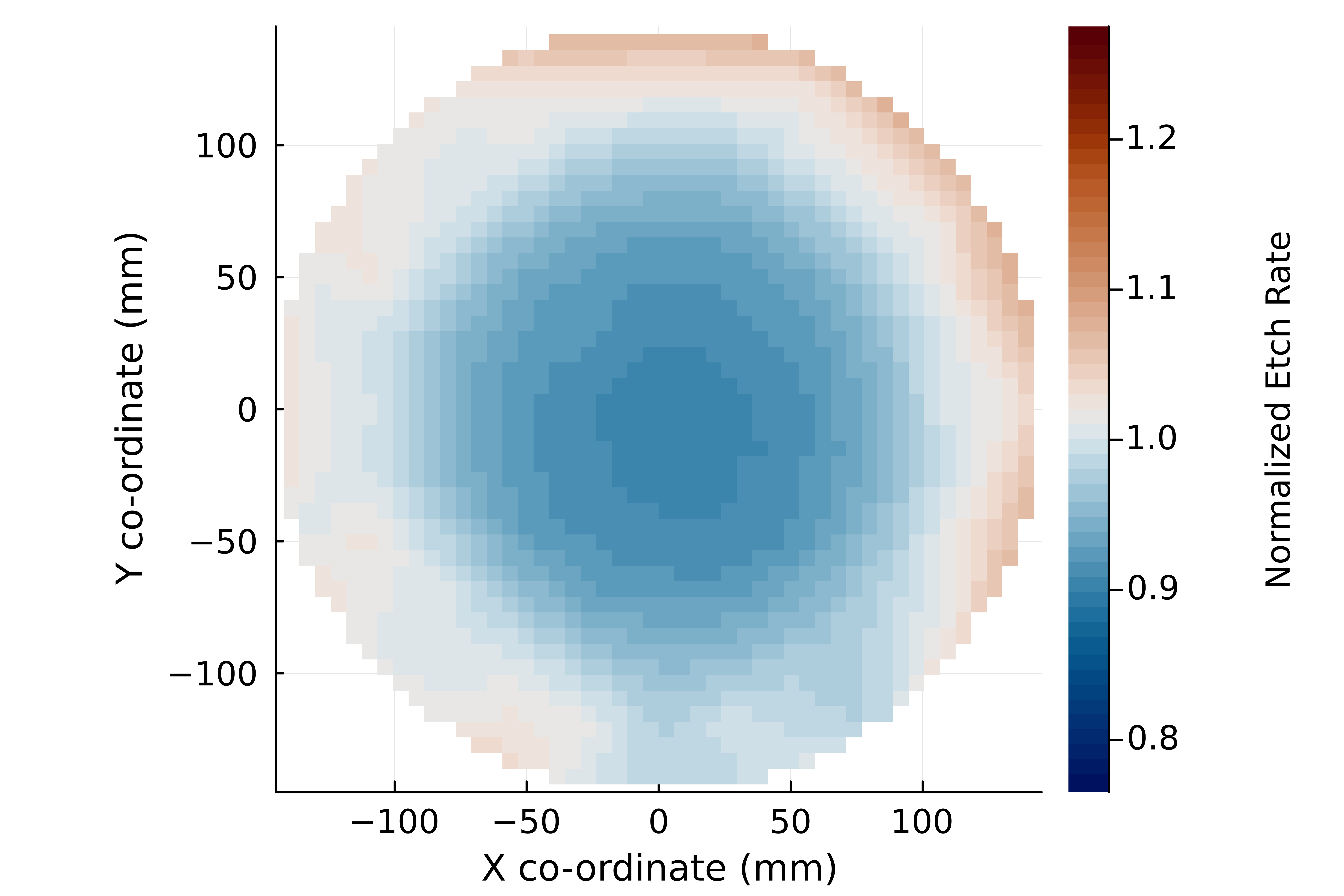}
    \caption{ }
    \end{subfigure}
~
    \begin{subfigure}[t!]{0.3\linewidth}
    \centering
    \includegraphics[width=\linewidth,  trim = {0in 0in 0in 0in}, clip]{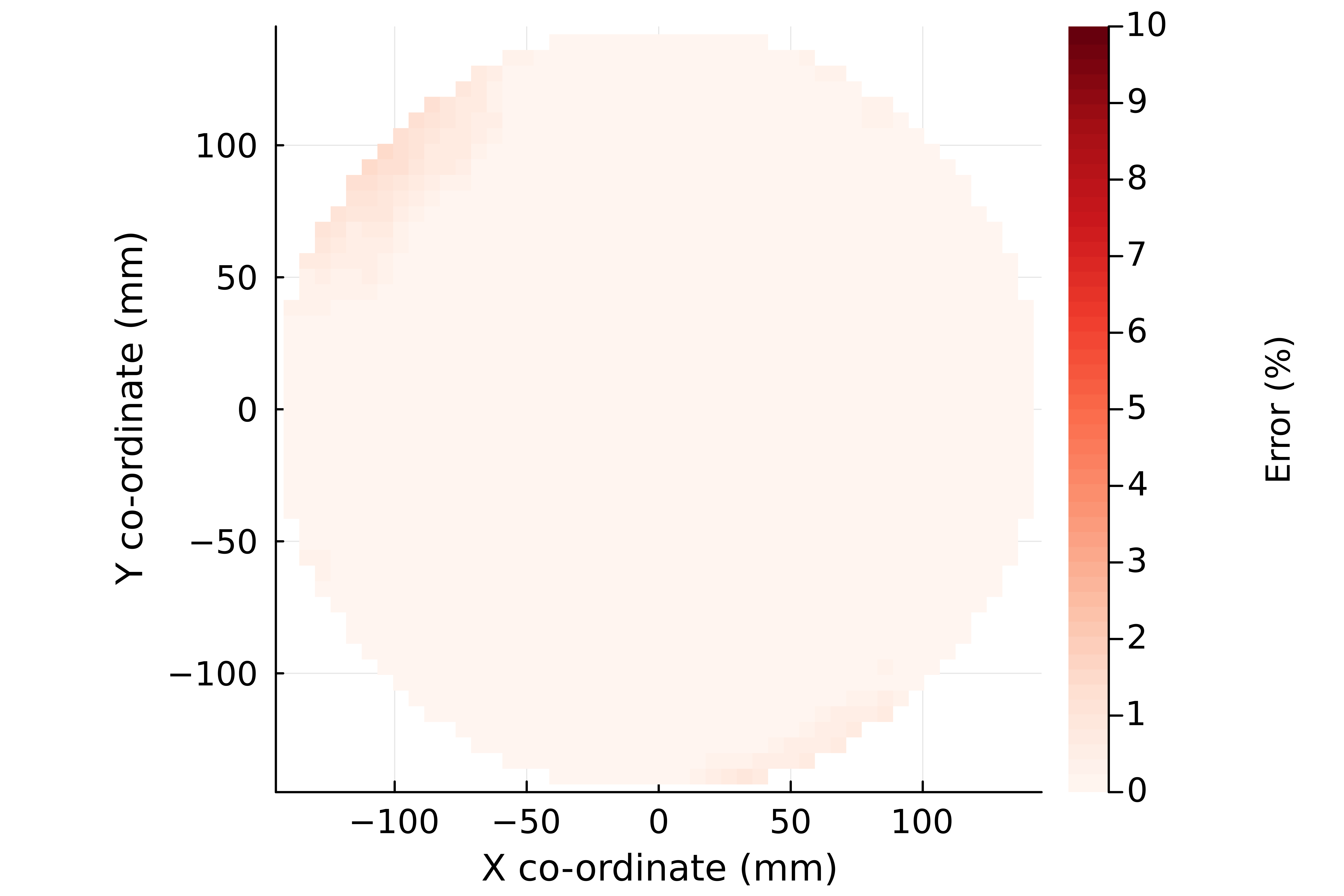}
    \caption{ }
    \end{subfigure}

    \caption{(a,b,c) The measured etch rate heatmap, the sensor-predicted etch rate heatmap and the local error $\mathcal{E}_{r,i}$  for the case where [$\Delta \overset{\cdot}{m_1}$, $\Delta \overset{\cdot}{m_2}$, $\Delta P$] = [$12$, $0$, $0$] (d,e,f) the same for the case where [$\Delta \overset{\cdot}{m_1}$, $\Delta \overset{\cdot}{m_2}$, $\Delta P$] = [$-2$, $7$, $0$] (g,h,i) and for the case where  [$\Delta \overset{\cdot}{m_1}$, $\Delta \overset{\cdot}{m_2}$, $\Delta P$] = [$0$, $0$, $-6$] }
    \label{fig:sensor_predicted_error_comparison}
\end{figure*}

We then perform the latent space variable interpolation into the sensor space for all the training wafers, while excluding the test wafers. This constructs the full surrogate model, which is then evaluated using the sensor data for the test wafers. Fig. \ref{fig:sensor_predicted_error_comparison} shows the heatmaps of the measured and sensor-predicted etch rate heatmaps for three test wafers, as well as the percentage error of the predicted wafer etch rate. The model is able to replicate the features of the etch rate scalar field well, with the characteristic lower etching rate at the center of the wafer, and a ring of higher etch rate at the edge of the wafer. We observe that the first two cases, which exist outside the training sensor space and on the boundary of the training space respectively, have far worse error than the last case which exists entirely within the training space. Furthermore, we also observe a consistent uniformity in the error of the sensor-predicted etch rates for the three wafers - no single region on the wafer has a considerably higher error than any other region.

\begin{figure}
     \centering
     \includegraphics[width=0.88\linewidth]{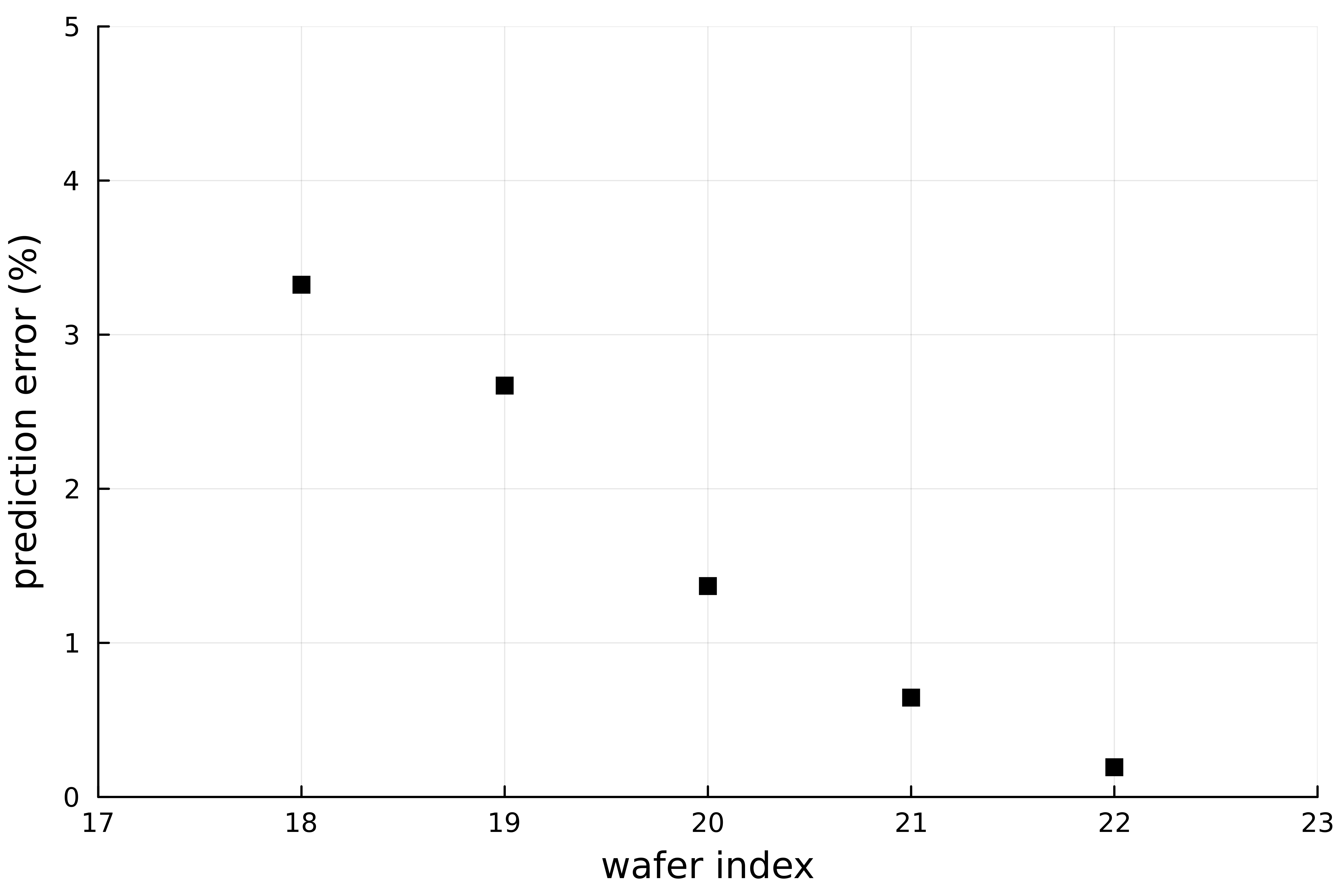}
     \caption{The sensor-predicted mean etch rate error $\mathcal{E}_{s,\mu}$ for the train and test wafers, evaluated at the measurement points.}
     \label{fig:sensor_predicted_error_all_wafers}
 \end{figure}

Fig. \ref{fig:sensor_predicted_error_all_wafers} shows the plot of the errors for the sensor-predicted etch rate means $\mathcal{E}_{s, \mu}$.  We observe that the error in the sensor-predicted mean is much higher than the reconstruction error.  Furthermore, the error for the extrapolation of the data as in the case of wafer 18 is higher than the error for the interpolation of the data as in the cases of wafers 19 through 22. We also observe that the error for wafers 21 and 22 is the least, as the input sensor data for these wafers exists entirely within the training space. Interestingly, the error for wafer 20 is much lower than the error for wafer 19. Observing the change in etch rate with the modification of each gas flow in fig. \ref{fig:etch_rate_AE} provides a possible explanation - the changes in the etch rate from changes in $\Delta \overset{\cdot}{m_2}$ are lesser than the changes in the etch rate from changes in $\Delta \overset{\cdot}{m_1}$. This implies that the etch rate prediction is more accurate for variation in $\Delta \overset{\cdot}{m_2}$ than in $\Delta \overset{\cdot}{m_1}$ as a result of the underlying chemistry of the etch process.

\begin{figure}
    \centering
    \includegraphics[width=0.88\linewidth]{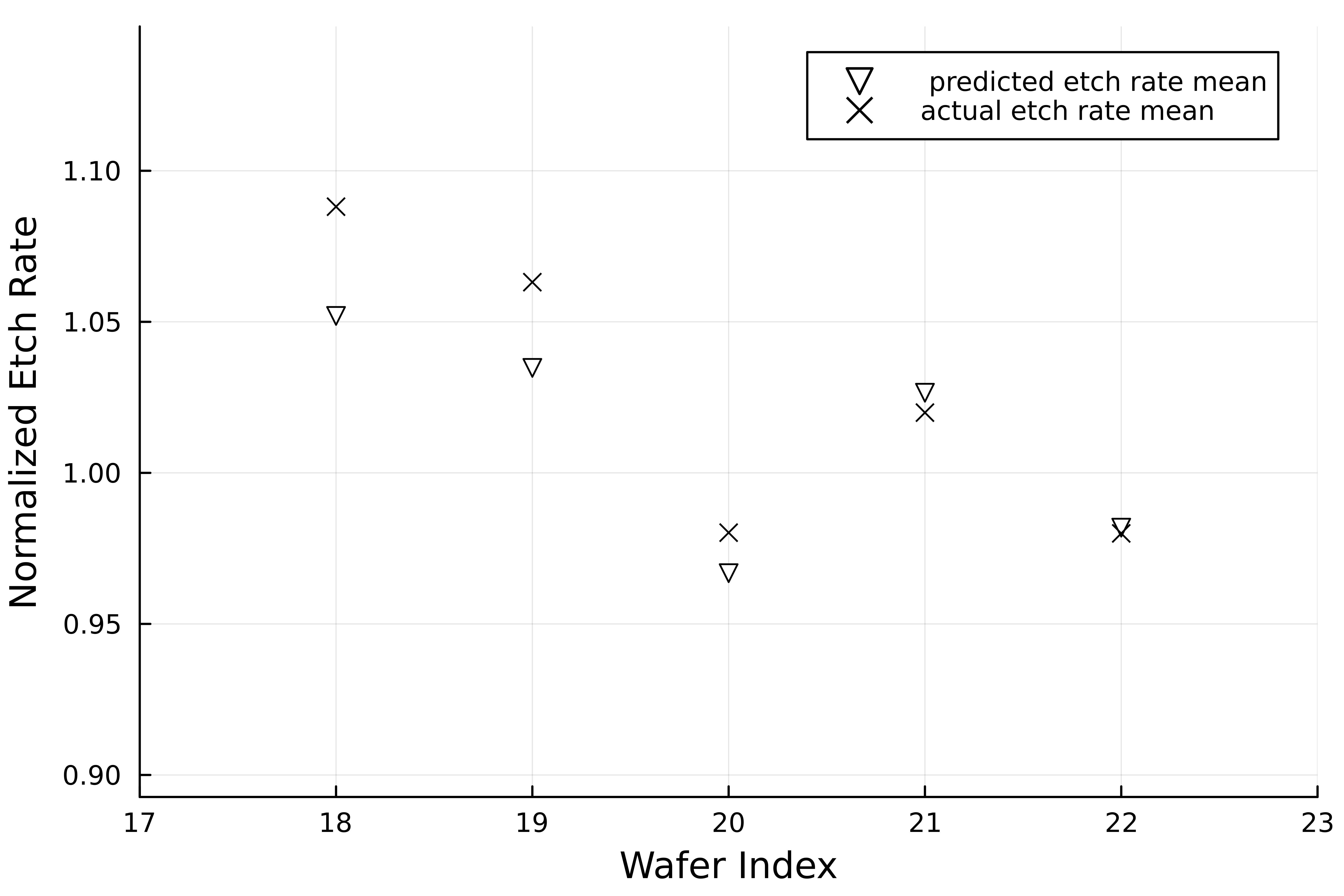}
    \caption{The sensor-predicted ($\mu(W^{r,u})$) and real mean etch rates ($\mu(W^{r,o})$) for the test wafers by the surrogate model.}
    \label{fig:etch_rate_sensor_predicted}
\end{figure}

The sensor-predicted etch rate means calculated at the measurement points $\mu(W^{r,s})$ are compared against the true etch rate means $\mu(W^{r,o})$ in fig. \ref{fig:etch_rate_sensor_predicted}. For the more uncertain predictions in the cases of wafers 18 through 21, the etch rate prediction is lower than the true etch rate. However, the opposite is true for wafers 21 and 22, which show a slight overprediction of the etch rate.

\section{Conclusions}

In the present work, we have developed and implemented a data-driven surrogate model for the prediction of the mean etch rates of a capacitively coupled reactive ion etcher from the tool's sensor data. The decoder half of an autoencoder trained on the etch rate measurements on wafers processed by the tool is used as a generative model to generate synthetic wafer profiles, with the value of the latent space vector for the synthetic profile being derived from interpolation maps created in sensor space. The reconstruction error of the trained autoencoder model is analyzed for both the train and test set, and is found to be less than one percent for all the considered cases. 

Training wafers are used to generate interpolation maps, with the sensor data from the test wafers used to generate etch rate heatmaps to test the model. The etch rate heatmaps and the mean etch rate are compared against the true data derived from metrology, with the model performing better when interpolating in between the data than extrapolating from the data. The etch rate means are found to be underpredicted by the model. However, the predicted etch rate means of the model estimate the true etch rate means to within a 5 percent error for all test cases.

\bibliographystyle{IEEEtran}
\bibliography{bibliography}

\newpage
 
\vspace{11pt}

\vspace{11pt}

\vfill

\end{document}